\documentclass{aa}
\usepackage[dvips]{graphics}
\usepackage{latexsym}
\def\gtsima{$\; \buildrel > \over \sim \;$}
\def\ltsima{$\; \buildrel < \over \sim \;$}
\def\prosima{$\; \buildrel \propto \over \sim \;$}
\def\gsim{\lower.5ex\hbox{\gtsima}}
\def\lsim{\lower.5ex\hbox{\ltsima}}
\def\simgt{\lower.5ex\hbox{\gtsima}}
\def\simlt{\lower.5ex\hbox{\ltsima}}
\def\simpr{\lower.5ex\hbox{\prosima}}

\def\h1{$h^{-1}$}
\def\beq{\begin{equation}}
\def\eeq{\end{equation}}
\begin{document}
\title{
The K20 survey. IV. The redshift distribution of $K_s<20$ galaxies: a 
test of galaxy formation models\thanks{Based 
on observations made at the European Southern Observatory,
Paranal, Chile (ESO LP 164.O-0560).}}
\author{
	A. Cimatti \inst{1}
	\and
	L. Pozzetti \inst{2}
	\and
	M. Mignoli \inst{2}
	\and
	E. Daddi \inst{3}
	\and
	N. Menci \inst{4}
	\and
	F. Poli \inst{5}
	\and
	A. Fontana \inst{4}
	\and
	A. Renzini \inst{3}
	\and
	G. Zamorani \inst{2}
	\and
	T. Broadhurst \inst{6}
	\and
	S. Cristiani \inst{7}
	\and
	S. D'Odorico \inst{3}
	\and
	E. Giallongo \inst{4}
	\and
	R. Gilmozzi \inst{3}
}
\institute{ 
INAF, Osservatorio Astrofisico di Arcetri, Largo E. Fermi 5, I-50125 Firenze, Italy 
\and INAF, Osservatorio Astronomico di Bologna, via Ranzani 1, I-40127, Bologna, Italy
\and European Southern Observatory, Karl-Schwarzschild-Str. 2, D-85748, Garching, Germany
\and INAF, Osservatorio Astronomico di Roma, via Dell'Osservatorio 2, Monteporzio, 
Italy
\and Dipartimento di Astronomia, Universit\`a ``La Sapienza'', Roma, Italy
\and Racah Institute for Physics, The Hebrew University, Jerusalem, 91904, Israel
\and INAF, Osservatorio Astronomico di Trieste, Via G.B. Tiepolo 11,
I-34131, Trieste, Italy
}
\offprints{Andrea Cimatti, \email{cimatti@arcetri.astro.it}}
\date{Received ; accepted }
\abstract{We present the redshift distribution of a complete sample
of 480 galaxies with $K_s<20$ distributed over two independent
fields covering a total area of 52 arcmin$^2$. The redshift completeness is 
87\% and 98\% respectively with spectroscopic and high-quality and tested 
photometric redshifts. The redshift distribution of
field galaxies has a median redshift $z_{med}\sim 0.80$, with $\sim$
32\% and $\sim$9\% of galaxies at $z>1$ and $z>1.5$ respectively. 
A ``blind'' comparison is made
with the predictions of a set of the most recent $\Lambda$CDM hierarchical 
merging and pure luminosity evolution (PLE) models. The hierarchical 
merging models overpredict and underpredict the number of galaxies at 
low-$z$ and high-$z$ respectively, whereas the PLE models match the median 
redshift and the low-$z$ distribution, still being able to follow the 
high-$z$ tail of $N(z)$. We briefly discuss the implications of this 
comparison and the possible origins of the observed discrepancies. 
We make the redshift distribution publicly available.
}
\titlerunning{The redshift distribution of $Ks<20$ galaxies}
\authorrunning{A. Cimatti et al.}  \maketitle

\section{Introduction}

The mass assembly history of galaxies remains one of the critical
issues in observational cosmology: did galaxies reach their present
stellar mass only recently (say, at $z\lsim 1$) ? Or were most
(massive) galaxies already in place by $z\sim 1$ ?
Spectroscopic surveys of faint galaxies selected in the $K$-band
currently offer the best opportunity to answer these questions
(Broadhurst et al. 1992). The main advantages with respect to
optically selected samples include: the direct sensitivity to the
galaxy stellar mass rather than to the ongoing/recent star formation 
activity (Gavazzi et al. 1996; Madau et al. 1998), the smaller 
K-correction effects, and the minor influence of dust extinction.

In this framework, we have completed an optical and near-infrared
spectroscopic survey down to $K_s<20$ (dubbed ``K20 survey'')
using ESO VLT telescopes and instruments, with full survey details
being given in Cimatti et al. (2002b, hereafter Paper III; see also
{\tt http://www.arcetri.astro.it/$\sim$k20/}).  The K20 sample
includes 546 objects to $K_s \leq 20$ (Vega system) over two independent
fields (52 arcmin$^{2}$ in total), so to be less affected by the cosmic
variance. The spectroscopic redshift completeness is 94\% and 87\% 
for $K_s \leq 19$ and $K_s \leq 20$ respectively. This makes the K20 
sample the largest and most complete spectroscopic sample of galaxies 
with $K_s<20$ available to date (see Paper III; cf. Cowie et al. 1996;
Cohen et al. 1999). Moreover, a 98\% redshift
completeness is reached for the $K_s \leq 20$ sample when including the 
photometric redshifts obtained with the available deep $UBVRIzJK_s$ imaging 
for those objects without a spectroscopic redshift. If stars and broad-line
AGNs are excluded, the total number of galaxies with $K_s\leq20.0$ and
with redshifts is 480.

In two previous papers based on the K20 survey we showed that Extremely 
Red Objects (EROs, defined by $R-K_s>5$) are nearly equally populated 
by old passively evolving galaxies and by dusty star-forming systems
at $z \sim 1$ (Cimatti et al. 2002a, Paper I). The number of all 
(old+dusty) EROs is strongly underpredicted by hierarchical merging 
models (HMMs), whereas old EROs have a density consistent with PLE 
models for passive early-type galaxies (Paper I), and are strongly 
clustered as opposed to dusty EROs (Daddi et al. 2002, Paper II).

In this Letter, we present and discuss the observed redshift distribution, 
$N(z)$, of all the galaxies in the K20 sample, irrespective of their color, 
and compare it to the expectations for the case of PLE of galaxies, as well
as to the predictions of various HMM renditions. The currently favored 
cosmological model is adopted, i.e., $H_0=70$ km s$^{-1}$ Mpc$^{-1}$, 
$\Omega_m=0.3$ and $\Omega_{\Lambda}=0.7$. 

\section{Observations vs. model predictions}

The observed $N(z)$ for the 480 galaxies (417 with spectroscopic 
and 63 with photometric redshifts respectively) with $K_s \leq 20$
is shown in Fig. 1. The redshift distribution can be retrieved from {\tt 
http://www.arcetri.astro.it/~k20/releases}.
The spike at $z\sim 0.7$ is due to two clusters (or rich 
groups) of galaxies respectively at $0.665<z<0.672$ (23 galaxies) and 
$0.732<z<0.740$ (33 galaxies) (see Paper III). The median redshift 
of $N(z)$
is $z_{med}=0.737$ and $z_{med}=0.805$, respectively with and without 
the two clusters being included. Without the clusters, the fractions 
of galaxies at $z>1$ and $z>1.5$ are 138/424 (32.5\%) and 39/424 
(9.2\%) respectively. The high-$z$ tail extends beyond $z=2$. 
The contribution of objects with only a photometric redshift 
becomes relevant only for $z>1.5$. 
The fractional cumulative distributions displayed in Fig. 2-3 (bottom
panels) were obtained by removing the two clusters mentioned above in 
order to perform a meaningful comparison with the galaxy formation 
models which do not include clusters (PLE models) or are averaged over 
very large volumes, hence diluting the effects of redshift spikes
(HMMs). 

\begin{figure}[ht]
\resizebox{\hsize}{!}{\includegraphics{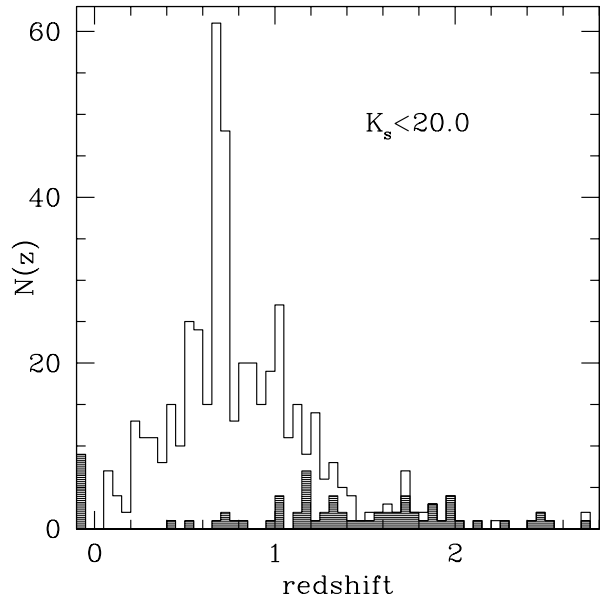}}
\caption{\footnotesize
The observed differential redshift distribution.
The shaded histogram shows the contribution of photometric redshifts. 
The bin at $z<0$ indicates the 9 objects without redshift. 
}
\label{fig:plot}
\end{figure}

\begin{figure}[ht]
\resizebox{\hsize}{!}{\includegraphics{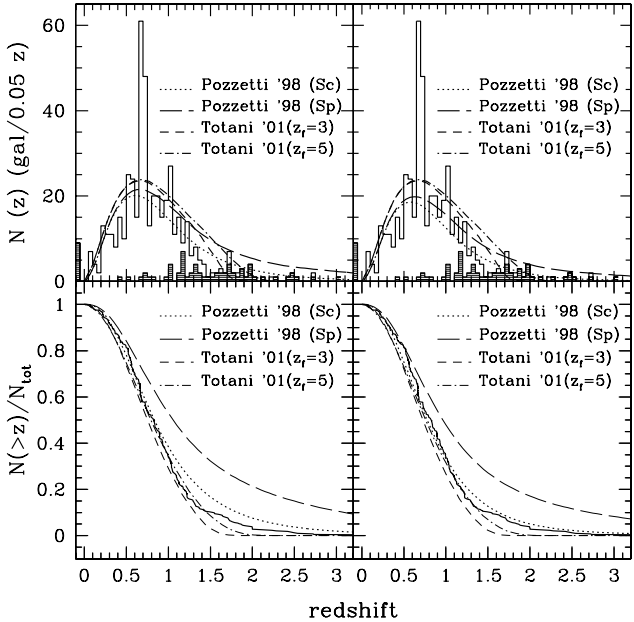}}
\caption{\footnotesize
{\it Top panels:} the observed differential $N(z)$ for 
$K_s<20$ (histogram) compared with the PLE model predictions. 
{\it Bottom panels:} the observed fractional cumulative redshift 
distribution (continuous line) compared with the same models.
The {\it left} and {\it right} panels show the models without and with the 
inclusion of the photometric selection effects respectively. Sc and Sp
indicate Scalo and Salpeter IMFs respectively.
}
\label{fig:plot}
\end{figure}

No best tuning of the models was attempted in this comparison, thus 
allowing an unbiased ``blind'' test with the K20 observational 
data. The model predicted $N(z)$ have been normalized to the K20 survey
sky area. We first discuss the case of PLE expectations, 
as derived by Pozzetti et al. (1996, 1998, PPLE hereafter), and 
Totani et al. (2001) (TPLE hereafter).

\subsection{Comparison with PLE models}

In the PPLE  model the present-day galaxy luminosity function is divided into 
5 Hubble types (E, S0, Sab-Sbc, Scd-Sdm, Im). The spectral evolution 
for each type is described by the Bruzual \& Charlot (1993) model 
(GISSEL version 2000) which reproduces the rest-frame colors and 
K-corrections of local galaxies. Exponentially declining star formation 
rate histories and solar metallicity are adopted. The age of each galaxy 
is set to 12.5 Gyr 
($z_f=5.7$) with the exception of Im galaxies (age=0.1 Gyr). The 
$e$-folding times are set to 0.3, 2, 10 Gyr, and $\infty$ for E, S0, 
Sab-Sbc, and Scd-Sdm-Im galaxies respectively. Dust extinction
is not taken into account. Besides the adopted 
cosmology, the only difference between the PPLE model used here and 
Pozzetti et al. (1996, 1998) is the use of the $K_s$-band local 
luminosity function from the 2MASS for different morphological 
types (Kochanek et al. 2001), which is in agreement with the overall 
local luminosity function of Cole et al. (2001). 
Fig. 2 shows the predictions of the PPLE models for two types of
initial mass function (IMF), Salpeter (1955) and Scalo (1986). With the 
flatter (i.e. Salpeter) IMF the intrinsic luminosity of both passively 
evolving and star forming galaxies increases more rapidly with redshift 
than in the case of the steeper (Scalo) IMF. As already discussed by
Pozzetti et al. (1996) (see also McCracken et al. 2000), the mild 
evolution allowed by the Scalo IMF is more consistent with the
observations of the rest-frame ultraviolet luminosity density up 
to $z\sim 1$ (e.g. Cowie et al. 1999), and reproduces most observables 
(galaxy counts, color and redshift distribution from the $U$ to the 
$K$ band) without invoking the strong number density evolution or dust 
extinction required by the Salpeter IMF (cf. Fig. 2). 

In the TPLE model galaxies are also divided into 5 types (E/S0, Sab,
Sbc, Scd, Sdm) and their spectral evolution is described using the
Arimoto \& Yoshii (1987) models. The $B$-band local luminosity function 
is used, and the Salpeter IMF and a top-heavy IMF with exponent 0.95 
are adopted for spirals and ellipticals respectively, with two options 
for formation redshifts: $z_f$=3 and $z_f$=5. Contrary to the PPLE model,
the evolution of metallicity and dust extinction in galaxies is treated 
in this model.

Fig. 2 shows fairly good agreement between the observed $N(z)$
distribution and the PLE models (with the exception of PPLE
with Salpeter IMF). The predicted and the observed
total number of galaxies with $K_s<20$ agree within 10\% for the
PPLE and the TPLE ($z_f=3$), 28\% for the TPLE ($z_f=5$), and 34\%
for PPLE with a Salpeter IMF. The predicted median redshifts are
just slightly higher than observed: $z_{med}$=0.83, 0.75 and
0.79 for the PPLE, TPLE ($z_f=3,\; 5$) models respectively, but
inconsistent with the PPLE with Salpeter IMF ($z_{med}$=1.05). This is
due to these PLE models somewhat overpredicting the number of galaxies
at $z\gsim 1.2$. However, as extensively discussed in Paper III,
because of the photometric selection effects present in the K20 sample
(partly due to the cosmological surface brightness dimming),
the total fluxes of spirals and ellipticals with $L$ 
\ltsima $L^*$ (i.e. the bulk of the K20 sample) are, on average,
underestimated by about 0.1 and 0.25 magnitudes, respectively. 
In order to assess the influence of such effects, we
compared the observed redshift distribution (down to our nominal
$K_s<20.0$) with the PPLE and TPLE models with $K_s<19.9$ for ``disk''
and $K_s<19.75$ for ``early-type'' galaxies. Fig. 2 (right panels) 
shows that when such selection effects are taken into account the PLE 
models become much
closer to the observed $N(z)$ thanks to the decrease of the predicted
high-$z$ tail. According to the Kolmogorov-Smirnov test, the PLE models are 
acceptable at 95\% confidence level, with the exception of the PPLE
model with Salpeter IMF (rejected at $>99$\% level).
We conclude that PLE models offer a satisfactory fit to the
observed $N(z)$ distribution, all the way to the highest redshifts in
our sample. 

\subsection{Comparison with hierarchical merging models}

For the comparison with the HMM predictions we were kindly granted
access to the model databases of Cole at al. (2000, C00 hereafter),
Somerville et al. (2001, S01 hereafter) and Menci et al. (2002, M02
hereafter). These models are tuned to reproduce some low-$z$
observable, such as the local galaxy luminosity function near $L\simeq
L^*$ (C00, M02) or the Tully-Fisher relation (S01). 
The main difference among the HMMs used here is the inclusion in 
S01 of the merging-promoted ``starburst'' mode of star formation besides 
the ``quiescent'' mode, the only one included in C00 and M02. The
starburst mode has the effect of increasing the overall star formation
at high redshift, when most of the merging takes place. Moreover,
merging between satellite galaxies within DM haloes is included in S01
and M02, but neglected in the C00, where satellites are allowed to
merge only onto the central massive galaxy. The M02 model without
merging between satellites is otherwise equivalent to the C00
rendition. The effect of merging between satellites is to deplete the
number of low-mass galaxies which aggregate to form larger units, thus
flattening the galaxy mass function at the faint end and slightly
increasing the number of intermediate mass galaxies (see S01 and M02 for
more details). All the HMMs used here adopt a Salpeter IMF.

The HMMs overpredict the total number of galaxies with $K_s<20$ in the 
K20 survey area by factors of 30-45\%. In particular, Fig. 3 (top panels) 
shows that all the HMMs show an excess of predicted galaxies at
$z<0.5$, e.g., by  a factor of $\sim 2.5$ at $z\sim 0.4$ for the C00
model, and $\sim 1.5-2$ for the M02 and S01 models, respectively. The
predicted median redshifts are $z_{med}$=0.59, 0.70 and 0.67 for the
C00, M02 and S01 models, respectively, thus being systematically lower
than the observed $z_{med}$. Moreover, all the HMMs have a deficit of
$z>1$ galaxies, in particular with a fraction at $z>1.5$ smaller by 
factors of $\sim$4 for the C00 model and $\sim$2$\div$3 for the M02 and S01
models. Such a discrepancy increases dramatically for higher redshifts,
where all the HMMs predict no galaxies with $z>2$. Fig. 3 (bottom
panels) illustrates that in the fractional cumulative distributions 
the discrepancy with observations appears 
systematic at all redshifts. The Kolmogorov-Smirnov test shows 
that all the HMMs are discrepant with the observations at $>99$\% 
level. The inclusion of the photometric biases exacerbates this 
discrepancy, as shown in Fig. 3 (right panels) for the M02 model 
calculated for $K_s<19.8$ in order to include an average photometric 
bias for spirals and ellipticals (the discrepancy for the C00 and 
S01 models becomes even stronger). 

\begin{figure}[ht]
\resizebox{\hsize}{!}{\includegraphics{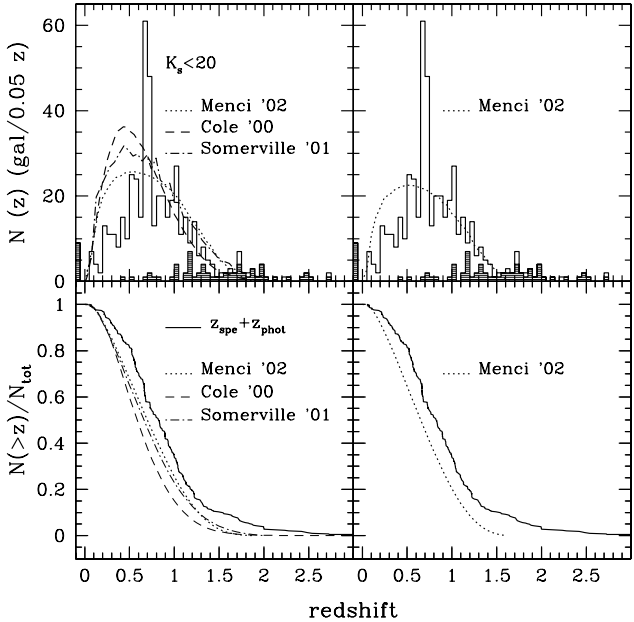}}
\caption{\footnotesize
{\it Top panels:} the observed differential redshift distribution for 
$K_s<20$ (histogram) compared with the HMM predictions. {\it Bottom
panels:} the observed fractional cumulative redshift distribution 
(continuous line) compared with the same models of top panels.
The {\it right} panels show the M02 model with the inclusion of the 
photometric selection effects.
}
\label{fig:plot}
\end{figure}

The excess of galaxies at $z\lsim 0.5$ seen in Fig. 3 is due to HMMs 
predicting too many low-mass, low-luminosity galaxies. This excess has 
typically afflicted HMMs, with the merging between satellites improvement 
being apparently insufficient to provide a better agreement with the
data. But in addition, HMMs underpredict the number of high-redshift 
objects. This is illustrated by Fig. 4, where the PPLE model is capable to 
reproduce the cumulative {\it number} distribution of galaxies at 
$1<z<3$ within 1-2$\sigma$, whereas the M02 model is always discrepant 
at \gtsima 3$\sigma$ level (up to $>5 \sigma$ for $1.5<z<2.5$).
All the results described in this section remain valid if 
shallower limiting magnitude thresholds are adopted (see Fig. 5).

\begin{figure}[ht]
\resizebox{\hsize}{!}{\includegraphics{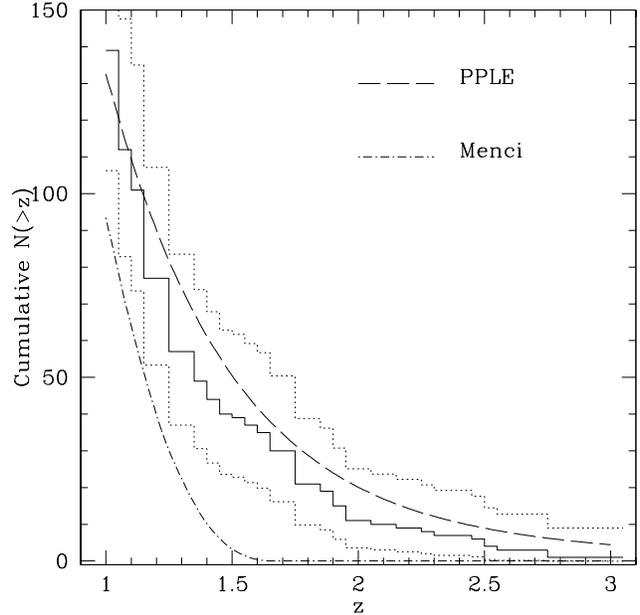}}
\caption{\footnotesize
The observed cumulative {\it number} of galaxies between $1<z<3$  
(continuous line) and the corresponding poissonian $\pm 3 \sigma$ 
confidence region (dotted lines). The PPLE (Scalo IMF)
and the M02 models are corrected for the photometric biases.
}
\label{fig:plot}
\end{figure}

\begin{figure}[ht]
\resizebox{\hsize}{!}{\includegraphics{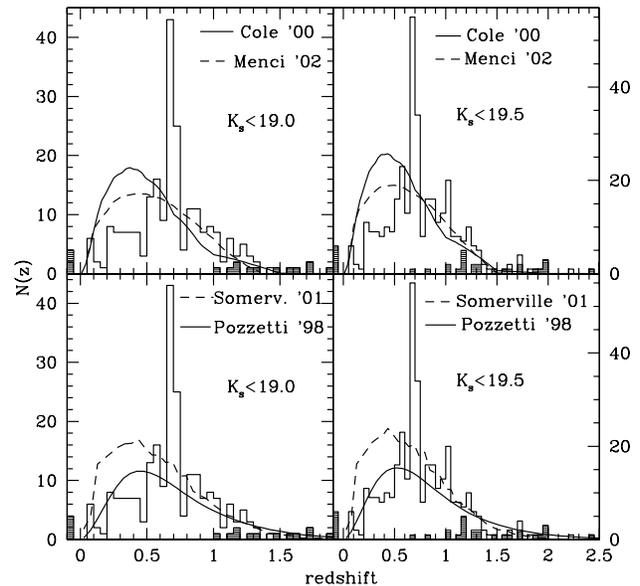}}
\caption{\footnotesize
The observed differential redshift distribution for $K_s<19$
and $K_s<19.5$ (histograms) compared with the model predictions
(not corrected for photometric selection effects). 
}
\label{fig:plot}
\end{figure}

\section{Discussion}

Early predictions of the expected fraction of galaxies at $z>1$ 
in a $K<20$ sample indicated respectively $\approx 60\%$ and $\approx 
10\%$ for a PLE case and for a (then) standard $\Omega_m=1$ CDM 
model (Kauffmann \& Charlot 1998). This version of PLE was 
then ruled out by Fontana et al. (1999). The more recent 
PLE and HMM used in this paper consistently show 
that for $z>1$ the difference between the predictions of different 
scenarios is much less extreme. These results partly from the now favored
$\Lambda$CDM cosmology which pushes most of the merging activity in
hierarchical models at earlier times compared to $\tau$CDM and SCDM
models with $\Omega_m=1$, and partly to different recipes for the star
formation modes, which tend to narrow the gap between HMMs and the PLE
case (e.g. Somerville et al. 2001; Firth et al. 2002).
The disagreement between the observed $N(z)$ and the predictions of 
the most updated HMMs based on a $\Lambda$CDM cosmology would then become 
even stronger in the case of old-fashioned CDM models with $\Omega_m=1$ 
because structures form later in a matter-dominated universe, and thus 
they would predict an even lower fraction of galaxies at high-$z$. 
In this respect, our results can be seen 
as additional evidence that the universe is not matter-dominated 
($\Omega_m<1$), and suggest that the HMMs may perform better if 
$\Omega_m$ is even lower than the currently favored $\Omega_m=0.3$. 

Nevertheless, the results of the K20 survey indicate that the shape
and the median of the observed redshift distribution of $K_s<20$
galaxies are in broad agreement with the expectations of PLE 
models, while disagree with the predictions of
current hierarchical merging models of galaxy formation. This
discrepancy refers to all galaxies, irrespective of color or morphology
selection, and therefore is more general than the already noted
discrepancies with EROs (Daddi et al. 2000; Paper I; Cimatti 2002). 
The poor performance of HMMs in accounting for the properties of even
$z=0\rightarrow\sim 1$ early-type galaxies has been 
emphasized in the past (e.g., Renzini 1999; Renzini \& Cimatti 1999).
Moreover, among low-redshift galaxies there appears to be a clear
anti-correlation of the specific star formation rate with galactic
mass (Gavazzi et al. 1996; Boselli et al. 2001), the most massive
galaxies being ``old'', the low-mass galaxies being instead dominated
by young stellar populations. This is just the opposite than expected
in the traditional hierarchical merging scenario, where the most
massive galaxies are the last to form.

On the other hand, the strong clustering of EROs seems to be
rather consistent with the predictions of CDM models of large scale 
structure evolution (Daddi et al. 2001, Paper II; Firth et al. 2002).
Thus, adopting the hierarchical merging $\Lambda$CDM scenario as 
the basic framework for structure and galaxy formation, the observed 
discrepancies may be ascribed to the heuristic algorithms adopted for
the star formation processes and their feedback, both within individual 
galaxies and in their environment. Our results suggest that HMMs should 
have galaxy formation in a CDM dominated universe to closely mimic the 
old-fashioned {\it monolithic collapse} scenario.
This requires to enhance merging and star formation in massive haloes 
at high redshift (say, $z\gsim 3$), while in the meantime suppressing 
star formation in low-mass haloes. For instance, Granato et al. (2001)
suggested 
the strong UV radiation feedback from the AGN activity during the era 
of supermassive black hole formation to be responsible for the suppression 
of star formation in low-mass haloes, hence imprinting a ``anti-hierarchical''
behavior in the baryonic component. The same effect may well result from 
the feedback by the starburst activity itself (see also Ferguson \& Babul 1998).

In summary, the redshift distribution presented in this paper,
together with the space density, nature, and clustering properties of
the ERO population (Paper I, Paper II) and the redshift evolution of
the luminosity and stellar mass functions derived for the K20 sample
(Pozzetti et al. 2002, Fontana et al. 2002) provide a new set of 
observables on the galaxy population in the $z\sim 1-2$ universe, 
thus bridging the properties of $z\sim 0$ galaxies with those of 
Lyman-break and submm/mm-selected galaxies at $z$ \gtsima 2-3. While 
making a step towards the
fully empirical mapping of galaxy formation and evolution, this set of 
observables poses a new challenge for theoretical models to properly reproduce. 

\begin{acknowledgements}
We are in debt with Carlton Baugh, Rachel Somerville and Tomonori Totani 
for providing their ``blind'' model predictions. We thank 
the referee, Nathan Roche, for useful comments and the VLT support 
astronomers for their assistance during the observations. AC warmly 
thanks ESO (Garching) for the hospitality during his visits.
\end{acknowledgements}

\end{document}